\documentclass{article}
\usepackage[utf8]{inputenc}

\usepackage{amsmath}
\usepackage{amsfonts}
\usepackage{caption}
\usepackage{pgffor}
\usepackage{amssymb}
\usepackage[pdftex,dvipsnames]{xcolor}
\usepackage{graphicx}
\usepackage{subcaption}
\usepackage{float}
\usepackage{blindtext}
\usepackage{csquotes}
\usepackage[colorlinks = true,
            linkcolor = blue,
            urlcolor  = blue,
            citecolor = blue,
            anchorcolor = blue]{hyperref}
\usepackage[capitalise,noabbrev]{cleveref}
\usepackage[margin=1in]{geometry}
\usepackage{enumerate} 
\usepackage{mathrsfs}

\title{Snowmass Topical Group Summary Report: IF07 - ASICs and Electronics}
\author{Gabriella Carini (Brookhaven National Laboratory),\and Mitch Newcomer (University of Pennsylvania), John Parsons (Columbia University) \and with contributions from the community}
\date{September 2022}

\usepackage[numbers,sort&compress]{natbib}
\usepackage{graphicx}

\usepackage[colorinlistoftodos,prependcaption,textsize=tiny]{todonotes}

\begin{document}

\maketitle

\section{IF07 Executive Summary}

In pursuit of its physics goals, the high energy physics community
develops increasingly complex detectors, with each subsequent
generation pushing to ever finer granularity. For example, the
emergence and growing dominance 
of particle-flow reconstruction methods has stressed
the increasing importance of very fine detector granularity, not
only for tracking detectors but also for calorimetry, and has also
initiated the push to "4D" detectors that combine precision measurements
of both spatial and time coordinates to their measurements of
energy and momentum. While these developments have led to enormous
growth in detector channel counts, the analog performance requirements 
for energy and momentum measurements, as well as spatial and timing precision, 
are being maintained or (more typically) even tightened.

These challenging detector requirements drive the corresponding development of
readout electronics. A variety of factors, including the growing
specialization of the functionality required, and the explosive
growth in channel counts and typically modest (if any) increases in the
material budget and the power and cooling budgets,
lead to the increasing reliance on custom-developed Application
Specific Integrated Circuits (ASICs). This trend is further 
exacerbated by challenges in the various experimental environments,
including radiation hardness requirements, cryogenic 
and deep-cryogenic operations, and 
space-based detector systems.
The development of custom ASICs in advanced technology nodes allows HEP detector
subsystems to achieve higher channel density, enhanced
performance, lower power consumption, lower mass, much greater radiation
tolerance, and improved cryogenic temperature performance than is possible
with commercial integrated circuits (ICs) or discrete components. 
The higher level of integration also leads to fewer components and
fewer connections, leading to higher reliability, as required by
experiments that can run for decades and that provide access to
the on-detector electronics at most annually, and sometimes never.

This writeup summarizes the work of ``IF07'', namely
Topical Working Group 7 of the
Instrumentation Frontier group of the Snowmass 2021 process. Group IF07
dealt with issues pertaining to ASICs and Readout Electronics. The
community efforts as part of IF07 were organized across 7 white papers
submitted to the Snowmass process. The first~\cite{whitepaper1}
discusses issues related to the need to maintain the talented
workforce required to successfully develop future HEP electronics
systems, and to provide the appropriate training and recruitment.
The remaining papers focused on electronics for particular detector
subsystems or technologies, including calorimetry~\cite{whitepaper2},detectors for fast timing~\cite{whitepaper4}, optical links~\cite{whitepaper5}, smart sensors using AI~\cite{whitepaper6},cryogenic readout~\cite{whitepaper7},
and RF readout systems~\cite{whitepaper8}. 

In the following section, a brief overview is provided of each 
of the white papers in turn.  
A white paper on silicon and photo-detectors was originally considered. This content is now captured by other frontiers with contributions form ASICs and readout electronics.

There are some overarching goals for advancing the field of readout and ASICs.  Efforts with broad impact include:
There are some overarching goals for advancing the field of readout and ASICs.  Efforts with broad impact include:
\begin{itemize}
\item[{\bf IF07-1}]: Provide baseline support and specialized {\bf training opportunities} for the instrumentation work force to keep the US electronics instrumentation and ASIC  workforce current.
\item[{\bf IF07-2}]: Improve mechanisms for {\bf shared access to advanced technology} providing broader access by the community to foster real exchange of information and accelerate development.
\item[{\bf IF07-3}]: Continue to develop methodologies to {\bf adapt the technology for operation in extreme environments}. Deep cryogenics, ultra-radio pure materials, radiation-harsh environments with limited power budget and long lifetime for all cases.
\item[{\bf IF07-4}]: Develop novel techniques to manage very high {\bf data rates}. Data reduction and optimization needs to be as close as possible to the generation point with acceptable power consumption.
\item[{\bf IF07-5}]: Create framework and platform for {\bf easy access to design tools}. Develop specialized online resources for the HEP community (e.g. system simulations and design repository). Provide the basis for true co-design R\&{}D efforts: from simulation to verification and implementation.
\end{itemize}

\section{Overview of IF07 White Papers}

A brief overview and high-level summary of each 
of the 7 IF07 white papers is provided below. For a more detailed
discussion of each of the various topics, the interested reader 
is referred to
the white papers themselves, and to the references therein.

\subsection{Workforce and Training Needs}
  
 Two  decades ago  as we embarked on the design of the LHC detector systems \emph{on detector}  readout was focused on the integration of custom sensors and multiple types of Application Specific Integrated Circuits with communications rates at or below 100 Mbps.   Significant effort went into the design of printed circuits on boards or flexible substrates. A new generation of ASICs were coming into use that could survive radiation tolerance levels consistent with the needs of LHC provided specialized layout techniques were employed. Designers were faced with learning new tools and rules for commercial ASICs that were attainable in a few months and system interfaces were being designed that allowed multiple institutions to design parts of a readout system nearly independently.   
 As LHC systems were being placed into service commercial ASICs and communications systems were leap frogging forward and significant change was moving the state of the art forward faster than our communities and budgets could maintain pace.   Today's  HL-LHC designs have been able to take advantage of advances that are still several technology generations behind the commercial state of the art to move what previously was multi-chip  functionality onto a single silicon substrate to make systems on a chip that operate at much higher clock rates and can hold and selectively readout more data  and contain most or all of the readout blocks for modules of thousands of small size sensors.   The price for successful submission of these far more complex, low power integrated circuits has been the requirement for a workforce with knowledge of a broad set of new tools for design and verification as well as formalized design management and integration tools that don't allow new versions or blocks to compromise the overall performance or design progress of these now highly complex systems on a chip. We also recognize the revolution in capabilities offered by new, complex FPGA's that comprise a large part of the off detector readout of detector systems designed over the past decade.  Here too the required knowledge base has expanded sufficiently far that coding is no longer an easy to skill to learn for this or next generation designers. We foresee the need for continuing workforce training past the qualifying academic degrees to update commercial design skills. Many designs require HEP specialized knowledge to enable successful first time designs for extreme temperature and radiation environments. Our community exploit the internet to provide an archive with searchable  access to design examples from previous generations of detectors to replace the institutional knowledge passed down from previous teams. This is  especially important given that the time between large detector system developments may exceed career lifetimes. In these extended interim's  it is recognized that core HEP instrumentation specialists need to  update their skills with practical projects to ensure their familiarity with the evolving state of the art in ASIC designs.   This can be encouraged by DOE provided annual  FOA's to support the design of service blocks that will be necessary in yet to be defined front end ASICs for next generation sensor arrays: High speed communication links, data storage blocks, PLL's,  power converters  and regulators  etc.   Having silicon tested designs for next generation ASICs will both speed up design cycles and help maintain workforce skill levels consistent with current (at the time of need) ASIC design tool familiarity to minimize the number of design submissions required to assure reliable ASIC performance.   It would also be helpful  for DOE to encourage a hybrid Instrumentation Based qualification for PHD degrees as part of a basic High Energy Physics degree.   
 In addition it will be beneficial to have instrumentation conferences with training available to introduce new design approaches or technologies. 
 We see the recent support from DOE for Instrumentation based traineeship to be an important step towards having a better trained, better informed  workforce not only for designing systems but also for the benefit of future peer reviewed systems.  
 
 \subsection{Calorimeter Readout Electronics}
 
 Calorimeters will continue to serve as key detector subsystems
 at future colliders, as well as in many other applications. 
 The traditional challenges of calorimeter readout electronics systems,
including providing high precision energy measurements over a very 
wide dynamic range, are increasingly compounded by the demands of much
finer granularity and correspondingly higher readout rates, as well
as the demand of providing precision time measurements in
the move toward ``4D calorimetry''. The on-detector location of
the frontend electronics, necessitated by signal-to-noise and
other requirements, imposes additional challenges, including
tolerance to radiation and/or magnetic fields, reliability over periods
of a year or more without maintenance, and power and cooling budgets.

Some of the key innovations in calorimetry that are driving ongoing and future readout electronics developments include 
particle flow algorithms, in which measurements of energies from 
calorimeters are combined with the momentum
measurements from charged-particle tracking detectors. 
and dual-readout detectors, which provide a more flexible combination
of the electromagnetic and hadronic components of a shower. Both
of these methods aim to significantly improve the energy
resolution, and both can be implemented
either with or without timing information as an added component. 

Particle flow algorithms are pushing calorimeter designs to
ever finer granularities, and therefore greatly increasing 
channel counts. As an example, the High Granularity Endcap 
Calorimeter (HGCAL)
being developed currently for the CMS HL-LHC upgrade
includes over 6 million
readout channels, a dramatic increase over the $\approx 200k$
channels of the ATLAS liquid argon (LAr) calorimeters that 
set the scale for "finely segmented" among the original LHC detectors.   

Meeting the challenges for frontend calorimeter readouts
has relied on custom ASICs for over 30 years, and ASICs will only
become increasingly important.  Fortunately, the higher level of
integration available today has permitted some consolidation;
for example, while the current ATLAS liquid argon (LAr) frontend
required development of 11 different custom ASICs, spread over
a variety of technologies, the HL-LHC development underway requires
only three. The higher level of integration is most clearly seen
in the digital realm, where for example the lpGBT chip in 65~nm CMOS
fulfills a number of functions, including clock and control
distribution, slow control monitoring, and data serialization
and formatting, that were spread over a number of different ASICs for
the original LHC developments. However, even in the analog realm some
consolidation has been achieved, such as the 130~nm ASIC developed for
LAr that combines the functions of both the preamplifiers and the shapers
from the original construction. 

Of key importance to maintaining the
ability to develop ASICs to meet the challenges of future calorimeters
will be to maintain affordable access to the specialized ASIC
processes in industry, and to qualify these processes concerning
their radiation tolerance and, in some applications, also their
performance in cryogenic environments. A fortunate development has
been the typically increasing radiation tolerance of new ASIC
processes with smaller feature sizes; while the original LHC readout
ASICs exploited a number of specialized processes, or used 
standard cells
and design rules that had to be explicitly developed to improve the
radiation tolerance, the HL-LHC developments can focus on
commercial
130~nm and 65~nm CMOS processes and use commercial standard cell
libraries.  The move to ever smaller feature-size ASIC processes
also greatly helps reduce the power consumption. However, the
corresponding evolution to lower power rails presents a significant
challenge for the very frontend analog circuits that must 
handle input signals over a very wide, often around 16-bit,
dynamic range.

Other challenges for the development of future calorimeter readouts
include powering systems, ranging from power supplies, to DC-DC
convertors and low dropout regulators (LDO) or on-chip
regulation, that can provide
the needed power in a practical and radiation-tolerant manner,
and high-speed optical links that can move off-detector the 
huge data volumes generated by the calorimeter frontends. Optical
links will be discussed further in Section~\ref{sec:links}. Powering
has long proved a thorny issue at the LHC, and carefully evaluating
over many years 
the radiation tolerance of commercial devices has clearly
demonstrated that the great majority would not survive the LHC 
conditions. As a result, industrial
partnerships were launched for both the original LHC and for the
HL-LHC to develop radiation-tolerant LDOs.

\subsection{Electronics for Fast Timing}
Picosecond-level timing will be an important component of the next generation of particle physics detectors. The ability to add a 4$^{th}$ dimension to our measurements will help address the increasing complexity of events at hadron colliders and provide new tools for precise tracking and calorimetry for all experiments. Time is crucial for background rejection in dark matter searches and neutrino detectors.  As resolution continues to increase, time will likely become an equal partner to position measurement in particle tracking and particle flow event fitting. All this has been enabled both by the rapid and continuous advance of fast electronics and the ability to generate fast signals with good signal/noise from a variety of solid state sensors, photodetectors, and micropattern gas-based detectors.

Fast silicon sensors with gain, (e.g. Low-gain Avalanche Detectors LGAD) and sensors without gain (e.g. 3D sensors), micro-pattern gaseous detectors, Cerenkov light and fast scintillators such as LYSO, microchannel plate (e.g. LAPPD), semiconductor-based photodetectors such as SPADs and SIPMs, and other sensors provide all very fast signals.
There are several R\&D efforts aimed at the development of fast ASIC electronics for future HEP applications with focus on using specific and advanced technology (e.g. SiGe, 28 and 22 nm CMOS) and implementing suitable concepts for the needed time resolution (e.g. full waveform digitation, TDC, monolithic solutions).

The design and optimization of the front-end amplifier is particularly important for fast electronics.  The front end typically defines the signal/noise and thus the time jitter of the system. It can also consume significant power. Each design must be optimized for its environment, considering signal source, input capacitance required resolution, and subsequent processing and developed as element of a larger system design effort.

Fast timing applications put special emphasis on noise and rise time, which often requires higher front-end power. Increased pixel density to cope with required resolution and occupancy although somewhat balanced by lower load capacitance, also strains the power budget. Improved per hit time resolution may require more complex processing of the input waveform with corrections for delta rays, nonuniform ionization and varying weighting fields. This will require more complex, power hungry on-chip calculations or increased waveform information sent to downstream processing.

For ultimate time resolution non-homogenous sensor responses must be compensated, either on-detector or as part of the processing chain. This is an opportunity to employ emerging technologies such as machine learning to front or back end systems to take advantage of all possible information.

Most of the sub-75 ps systems demonstrated to date have either been in small, well constrained systems, or in beam tests. Distribution and maintenance of the clock system will be a challenge for large systems. In large systems timing must be monitored and temperature and aging effects compensated. Optical transceivers can have several ps/degree delay variation. This has been considered in detail for Xilinx Ultrascale transceivers and techniques have been developed to provide 1 ps phase resolution.

Using a combination of these instrumentation techniques and developments, including Constant Fraction Discrimination (CFD), waveform sampling combined with precise clock distribution and newly developed sensors we expect that it will be reasonable for future fast timing  detector sub-systems to set a 10pS timing resolution  goal that will allow for  unprecedented accuracy and significantly improve the physics reach of next generation  high rate, collider detectors. We anticipate that the grand challenge of measuring particles with a
resolution of about 1 ps may be possible and would provide revolutionary physics opportunities.

\subsection{Optical Links}
%\label{sec:links}

The dramatic increase in channel count that has resulted
from detectors with ever finer granularity, plus the need to
deliver in real time either all or at least more of the full-granularity,
full-precision detector readout data, has placed increasing 
demands on the optical links used to transmit the data from
the on-detector frontend electronics to the off-detector
digital processing and TDAQ systems. The radiation-tolerance
specifications for the detector mounted links has led to reliance on custom
developments, for the most part, though the original LHC detectors
did use some commercial link components which were tested to be
sufficiently radiation hard. 

The radiation requirements plus the reliance on custom solutions have
resulted in link speeds per fiber which are significantly lower than
used in commercial systems. The original LHC detectors employed
on-detector optical links from several 100~Mbps up to 1.6~Gbps per fiber.
The per-fiber bandwidth increased
to 5~Gbps for the recently completed Phase I upgrade of the LHC experiments, 
and currently to 10~Gbps for the ongoing HL-LHC developments. This substantial increase in
per-fiber bandwidth has, however, not kept pace with the growth in
data volume. For example, the original ATLAS LAr readout used a
single 1.6~Gbps link per 128 channels, while the corresponding HL-LHC readout
will need 22 fibers at 10~Gbps each for 128 channels. The
relatively low (up to 10~Gbps) per-fiber bandwidth, compared to industry rates of
up to 56~Gbps, results
in large fiber plants and an inefficient use of the SerDes resources
of the fast (and expensive) FPGAs that are typically used to
receive and process the on-detector digital data once it arrives off-detector and
away from radiation.

Realization of a functional optical link requires several building 
blocks, including a data fan-in and serializer, combined with a
electrical-to-optical conversion coupled to an optical module that
couples to the fibers. 

For serializers, the custom-designed 5~Gbps GBT was developed by CERN in 128~nm 
CMOS, predominantly for use in the Phase I upgrades of the LHC detectors. 
Following this successful mode, the 10~Gbps lpGBT was
then developed in 65~nm CMOS, and will be used in most HL-LHC on-detector electronic
readout applications.
R\&D is underway that aims to increase the bandwidth by a factor of two
in the short term, still within 65~nm CMOS, and then to move
to 56~Gbps utilizing 28~nm CMOS.

The critical components for the optical modules themselves
have been mostly
satisfied by commercially available VCSELs and pin diodes, which
must be painstakingly selected for radiation tolerance. However, a custom
optical module that integrates these functions plus the optical
connections is still required, due mostly to the tight spatial and mechanical
constraints, as well as material budget, imposed, in particular, 
by the inner detectors at the LHC and HL-LHC.

It is apparent that future detector readouts will continue to deliver increasingly
large data volumes. Meeting the corresponding readout bandwidth requirements, 
and in particular for on-detector environments with a significant 
radiation-tolerance requirement, poses a number of  
very significant challenges.  Meeting 
these needs will require ongoing R\&D, to facilitate the continued evolution
to higher bandwidths, and in particular per-fiber bandwidths.

\subsection{Smart Sensors Using Artificial Intelligence}

Modern particle physics experiments and accelerators, exploring nature at increasingly finer spatial and temporal scales in extreme environments, create massive amounts of data which require real-time data reduction as close to the data source and sensors as possible. The demand for increasingly higher sensitivity in experiments, along with advances in the design of state-of-the-art sensing systems, has resulted in rapidly growing big data pipelines such that transmission of acquired data for offline processing via conventional methods is no longer feasible. Data transmission is commonly much less efficient than data processing. Therefore, placing data compression, extracting waveform features and processing as close as possible to data creation while maintaining physics performance is a crucial task in modern physics experiments.  The implementation of Artificial intelligence (AI) and machine learning (ML) in near-detector electronics is a natural path to add capability for detector readout. It will enable more powerful data compression and filtering which better preserves the physics content of experiments, reduces downstream system complexity, and provides fast feedback and control loops. While the application of AI/ML is growing rapidly in science and industry, the needs of particle physics for speed, throughput, fidelity, interpretability, and reliability in extreme environments require advancing state-of-the-art technology in use-cases that go far beyond industrial and commercial applications.

AI, and more specifically ML, has recently been demonstrated to be a powerful tool for data compression, waveform processing, and analysis in physics and many other domains. While progress has been made towards generic real-time processing through inference including boosted decision trees and neural networks (NNs) using FPGAs (Field Programmable Gate Arrays) in off-detector electronics, ML methods are not commonly used to address the significant bottleneck in the transport of data from front-end ASICs to back-end FPGAs. 
Embedding ML as close as possible to the data source has a number of potential benefits
\begin{itemize}
    \item ML algorithms can enable powerful and efficient non-linear data reduction or feature extraction techniques, beyond simple summing and thresholding, which better preserves the physics content that would otherwise be lost;  
    \item This could in turn reduce the complexity of down stream processing systems which would then have to aggregate less overall information all the way to offline computing; 
    \item This enables real-time data filtering and triggering like at the LHC and the EIC which would otherwise not be possible or be much less efficient; or in the case of cryogenic systems, creates less data bandwidth from cold to warm electronics and thus reduce the system complexity;
    \item Furthermore, intelligent processing as close as possible to the source will enable faster feedback loops. For example, in continuous learning applications, if the data is part of a control or operations loop where feedback is needed such as in quantum information systems or particle accelerators. 
\end{itemize}

With rapidly growing machine learning applications comes the acute need for their efficient hardware implementations. Most of the efforts are focused on digital CMOS technology, such as implementations based on general-purpose TPUs/GPUs, FPGAs, and more specialized ML hardware accelerators.  The steady improvements in such hardware platforms' performance and energy efficiency over the past decade are attributed to the use of very advanced, sub-10-nm CMOS processes and holistic optimization of circuits, architectures, and algorithms. 
The opportunities for building more efficient hardware may come from biological neural networks. Indeed, it is believed that the human brain, with its $>$1000$\times$ more synapses than the weights in the largest transformer network, is extremely energy efficient, which serves as a general motivation for developing neuromorphic hardware. There is a long history of CMOS neuromorphic circuits. However, unleashing the full potential of neuromorphic computing might require novel, beyond-CMOS device and circuit technologies that allow for more efficient implementations of various functionalities of biological neural systems. 

There are several ongoing R\&D efforts in our community focused on on-detector AI/ML and the key elements of both the design and implementation, and the design tools themselves.  Though early in the exploration of these applications, the existing work highlights the needs for configurable and adaptable designs and open-source and accessible design tools to implement efficient hardware AI. Novel ML techniques are particularly important for resource-constrained real-time AI; advancing design, implementation, and verification tools which accelerate the development process; and emerging microelectronics technologies which could provide large gains in efficiency and speed. Different classes of future applications require dedicated investments to advance our capabilities in this field. 

Many emerging devices and circuit technologies are currently being explored for neuromorphic hardware implementations. 
Neuromorphic inference accelerators utilizing analog in-memory computing based on floating gate memories are perhaps the closest to widespread adoption, given the maturity of such technology, the practicality of its applications, and competitive performance with almost 1000x improvement in power as compared to conventional (digital CMOS) circuit implementations. The radiation hardness of these techniques and their applicability for robust performance in extreme environments is yet to be evaluated.

\subsection{Cryogenics Readout}

Many applications, including direct Dark Matter detection
and neutrino experiments, require electronics
designed to perform within cryogenic environments.  

Some operate at deep cryogenic levels going from  liquid helium temperatures at a few Kelvin down
to milli-Kelvin. 
Time projection chambers (TPCs) used in neutrino detection and dark matter searches  use noble liquids as a target material.  Most current generation  TPC's keep the active electronics in the warm, outside the cryostat. 

The sheer scale of the DUNE LArm TPC  being designed today makes it impracticalto
bring all the raw analog detector signals into the warm, and instead  embeds the active front end
electronics in the cold. For the DUNE wire-based LAr TPC readout, the required cold functionality is achieved by a set of 3 custom ASICs, one that amplifies and provides analog filtering,
one that digitizes at 2~MSPS, and one that collects and serializes the data for transmission over copper cables to the warm electronics outside the cryostat.  
 
The DUNE near detector will use a low power pixel readout utilizing the self-triggering LarPix 64-channel pixel ASIC submerged in  LAr.  Four programmable readout paths on the chip allow a board-level fault tolerant readout path. Track reconstruction results from prototype boards populated with 100 ASICs each have shown very promising physics results. 
 
A second very low power readout is being investigated by the Qpix Collaboration that utilizes a novel approach to record the times of arrival of a fixed unit of charge, enabling offline reconstruction of the ionization current sensed at the pixel with a sub-femto Coulomb unit charge sensitivity.

Deep-cryogenic CMOS circuits and systems have seen rapid development in recent years, with quantum computing being the primary technological and economic driver. Early work focused on studying and modeling the properties of devices, including MOSFETs, resistors, and capacitors, in various CMOS technologies at cryogenic temperatures. The results suggested that devices in modern nm-scale fabrication technologies (including bulk CMOS, silicon-on-insulator, and FinFET) function successfully at temperatures as low as 50 mK. The main deleterious effects of cryogenic operation on nanoscale MOSFETs include a moderate increase in threshold voltage and degradation in 1/f noise and matching properties, both of which can be overcome using well-known precision circuit design techniques. A potentially more serious problem is the self-heating of such devices due to internal power dissipation. For example, a recent study shows that the channel temperature of 40~nm bulk CMOS transistors can increase by over 40~K compared to a LHe ambient of 4.2 K when dissipating only 2 mW. Thus, integrated electrothermal modeling is an important requirement for successfully designing deep-cryogenic CMOS systems.
Given the availability of reliable cryogenic CMOS device models, research has focused on using these models to develop and test functional blocks necessary for qubit control and readout, including cryogenic frequency synthesizers, low-noise amplifiers, and circulators. Cryogenic operation of embedded memory, including standard SRAM cells, has also been demonstrated. The availability of these building blocks has motivated the recent development of complete cryogenic CMOS qubit control and readout interface and monolithic quantum processors that integrate on-chip silicon charge qubits with CMOS readout electronics.
Precision sensing is another major driver for cryogenic CMOS circuits and systems where ultra-low-noise analog front ends is a major application area.

\subsection{RF Electronics}
RF electronics utilizes  the  RF domain to sense or control EM radiation well below  ionization energies to make highly sensitive measurements of natural phenomena.  The sophistication and sensitivity of RF related detection systems has improved dramatically with advances in wireless communications.  HEP now has the opportunity to probe electromagnetic field signals measuring amplitude, phase, frequency, and polarization  with unprecedented signal-to-noise and orders-of-magnitude improvement in signal processing throughput per unit of power consumption. 
Combining  RF A/D, D/A  high precision and bandwidth performance and High speed FPGA's to readout arrayed superconducting sensors such as: Transition Edge Sensors TES, Kinetic Induction Detectors KID and superconducting bolometer   systems with reduced the sensor size and will give us a better view of the universe. 
 
Measurements of the extreme red shifted 21\,cm (1420 MHz) hyperfine transition of neutral atomic hydrogen are helping researchers to explore the early Universe searching in the red shifted range of  10-500 MHz  inferring  times before the creation of stars where a homogenous mix of materials allow robust theoretical predictions to be tested. The IF7 RF white paper describes the sophisticated very low noise techniques that are required to achieve the necessary electronics noise levels.  To achieve reasonable levels of detection, undistorted signal backgrounds can only be found locally on the far side of the moon. Multiple proposals are being developed with the objective of installing sensitive antennae and electronics there with lunar orbiting satellite relay stations.   The results of these explorations will benefit the ongoing development of cosmological models and may be an indicator of new physics.

\newpage
\bibliographystyle{JHEP}

%\bibliography{bibliography} 

%\section{References}
%\bibliographystyle{plain}
%\bibliography{references}
\end{document}